\documentstyle[11pt,newpasp,twoside,epsfig]{article}

\index{supernova remnants}
\index{molecular clouds}
\index{G349.7+0.2}
\index{infrared}
\index{masers}
\index{shock waves}

\def\edcomment#1{\iffalse\marginpar{\raggedright\sl#1\/}\else\relax\fi}
\reversemarginpar

\newcommand{\e}[1]{$10^{#1}$}
\newcommand{\ee}[1]{$\times 10^{#1}$}         
\newcommand{\kms}{~km\,s$^{-1}$} 
\newcommand{\cm}[1]{~cm$^{#1}$}
\newcommand{\ergs}{~erg\,s$^{-1}$\,cm$^{-2}$\,sr$^{-1}$}
\newcommand{\jb}{\,Jy\,beam$^{-1}$}

\newcommand{\co}{$^{12}$CO}    
\newcommand{\tco}{$^{13}$CO}

\newcommand{\hcop}{HCO$^+$}
\newcommand{\oh}{OH(1720 MHz)\ }
\newcommand{\h}{H$_2$}
\newcommand{\candy}{G349.2+0.7}

\begin{document}
\title{Molecular Diagnostics of Supernova Remnant Shocks}
 \author{J. S. Lazendic\altaffilmark{1}, M. Wardle, A. J. Green}
\affil{School of Physics, University of Sydney, NSW 2006, Australia}
\author{J. B. Whiteoak}
\affil{Australia Telescope National Facility, CSIRO, PO Box 76, Epping NSW 1710, Australia}
\author{M. G. Burton}
\affil{School of Physics, University of New South Wales, Sydney NSW 2052, Australia}

\altaffiltext{1}{Current address: Harvard--Smithsonian Center for Astrophysics, 60 Garden Street, Cambridge, MA 02138}

\begin{abstract}

We have undertaken a study of radio and infrared molecular--line
emission towards several SNRs in order to investigate molecular
signatures of SNR shocks, and to test models for \oh maser production in SNRs.  Here we present results on G349.7+0.2.

\end{abstract}

\section{Introduction}

Shocks driven by a supernova remnant (SNR) into a molecular cloud
compress, accelerate and heat the gas, and can contribute to disruption of
the molecular cloud. Despite the close association between SNRs and
molecular clouds (Huang \& Thaddeus 1986), establishing unambiguous cases
of interaction has been difficult, mostly because of confusion resulting
from unrelated clouds along the line of sight to the SNR and the
possibility of chance alignment. Studies of SNR/molecular cloud
interactions have been re--energized by the recent discovery of an
association between \oh maser emission and SNRs (see Koralesky et
al.~1998, and references therein). This maser emission requires very
specific conditions: gas density of $\sim$\e{5}\cm{-3}, gas temperature of
50--125 K, and OH column density of $\sim$\e{16}--\e{17}\cm{-2} (Lockett,
Gauthier \& Elitzur 1999). These conditions can be satisfied in the
cooling gas behind a non--dissociative C--shock, irradiated by the X--ray
flux from the SNR interior (Wardle 1999). The presence of \oh masers in
SNRs provides two crucial pieces of information for studying SNR shocks in
molecular clouds: 1) the velocity of a cloud associated with an SNR, and
2) the precise location of the on--going interaction.

\begin{figure}
\centerline{\epsfig{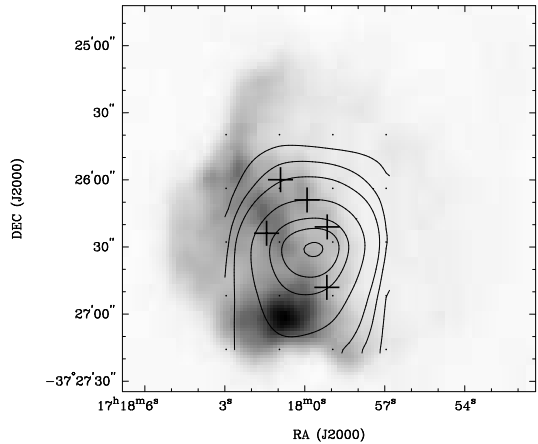}}
\caption{G349.7+0.2 is located at a distance of $\sim$23 kpc, and
is the third brightest galactic SNR at radio wavelengths. Five OH(1720
MHz) masers (marked with crosses) have been detected along the
bright emission ridge of the SNR (Frail et al.~1996). The \co\ 1--0
emission obtained with SEST is shown, integrated between 10 and
20\kms\  (contours), and overlaid on the 18--cm radio continuum greyscale
image obtained with ATCA with resolution of 9$\arcsec\times5\arcsec$. The contour levels are: 16, 24, 32, 48, 63, 71, 78 K\kms. The dots mark the grid positions of the SEST observations.}
\label{fig-candy-co}
\end{figure}

\section{Molecular gas towards the OH/SNR sites}

For millimeter--line observations of molecular clouds interacting with
SNRs, we used the Australia Telescope Mopra antenna and the Swedish--ESO
Submillimetre Telescope (SEST). These observations had sufficient (sub--km/s)
velocity resolution to associate the molecular
gas with the OH masers, but suffered from poor spatial resolution
(20\arcsec--50\arcsec\ for sources more than 8.5 kpc
away). Observations of species such as \co, \tco, CS,
and \hcop\  were used to constrain the physical parameters of
the ambient and shocked molecular gas. The observed molecular--line
profiles imply the presence of a C--type shock in ambient gas with density
of $\sim$ \e{4}\cm{-3} and temperature of $\sim$ 20 K. The shocked gas has a
density of \e{5}--\e{6}\cm{-3} and temperature of $\ga$ 40 K. The
latter properties of the molecular gas  are consistent with the
requirements for \oh maser production.

\section{\h\ emission}

\begin{figure}
\centerline{\epsfig{file=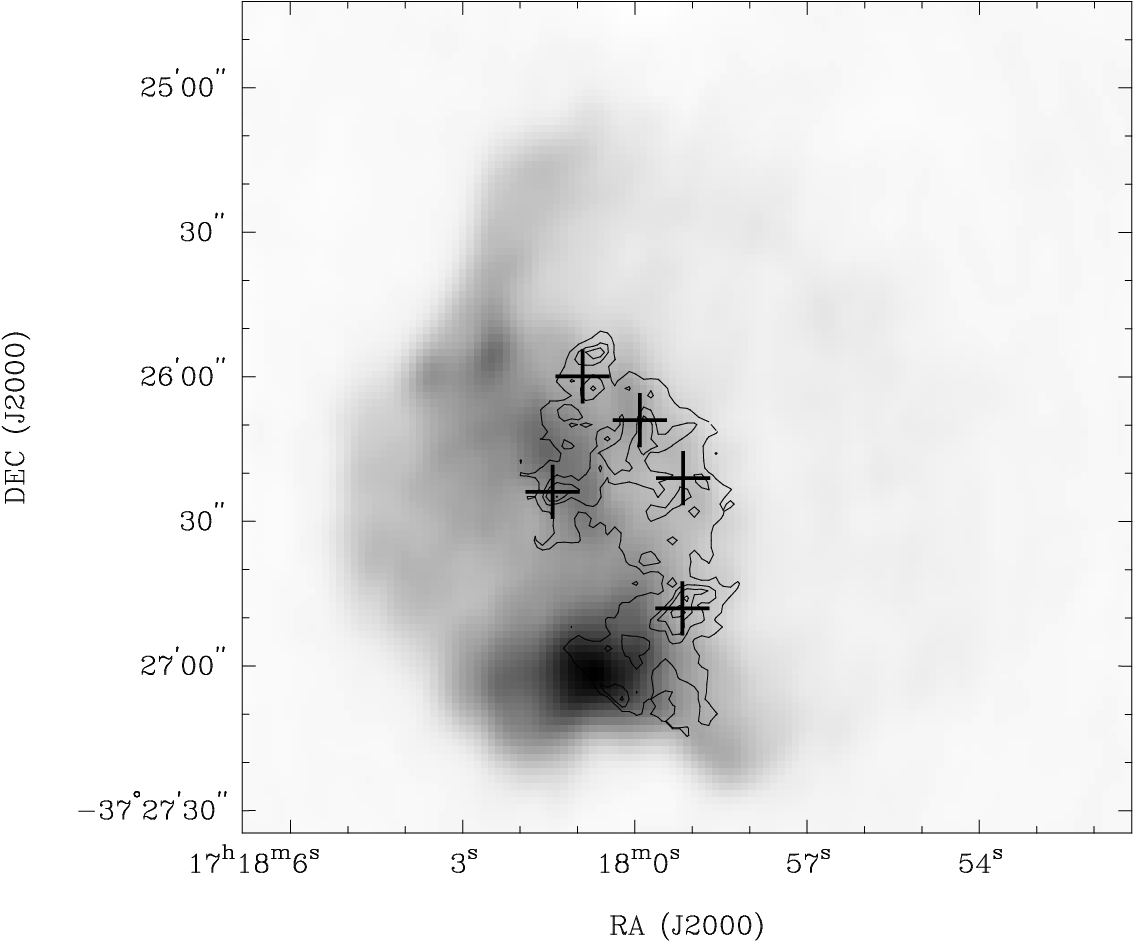,width=8cm}}
\centerline{\epsfig{file=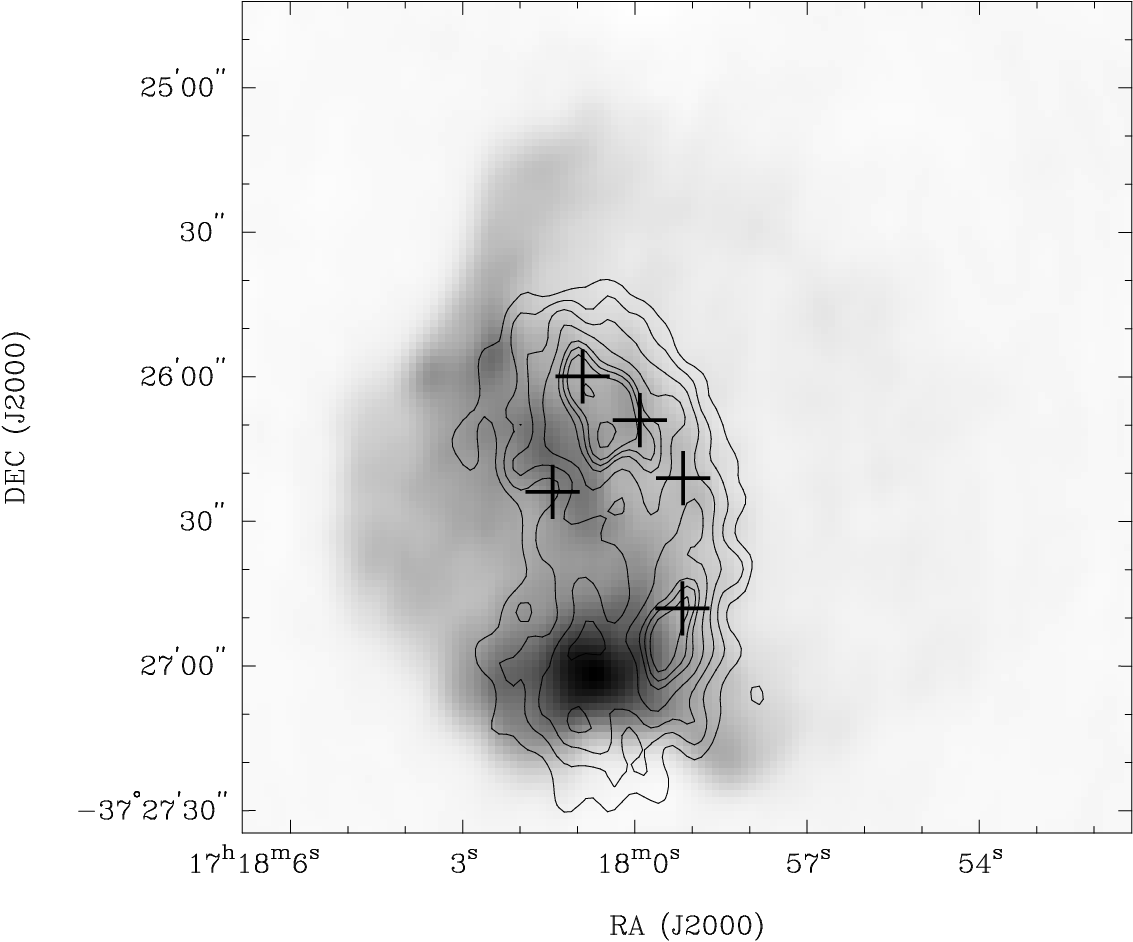,width=8cm}}
\caption{({\it Upper}) Shock--excited \h\ emission was detected
towards the western edge of \candy. The distribution of
2.12$\mu$m \h\ 1--0 S(1) emission (contours) is superimposed on the
18--cm ATCA image of G349.7+0.2. The contours are: 3.2, 9.7, 16.1 and
25.7 \ee{-5}\ergs.  \h\ observations revealed highly clumped molecular
gas. An excellent correlation between the \h\ clumps and \oh maser
positions confirms the shock excitation of the masers.
({\it Lower})  OH absorption towards \candy\ was found at the radial
velocity of the OH masers of $\sim$+16\kms. The map of OH
line--to--continuum ratio is shown, integrated between 10 and 20\kms\
(contours) overlaid on the 18--cm continuum image. The OH contours range from $-$0.01 to $-$0.03 \jb\ (beam is 15\arcsec$\times$12\arcsec).}
\label{fig-candy}
\end{figure}

Millimeter--line observations were complemented by near--infrared
(NIR) observations of \h\ emission with the University of New South
Wales Infrared Fabry--Perot (UNSWIRF) narrow--band imaging
spectrometer (Ryder et al.~1998), on the Anglo--Australian Telescope.
UNSWIRF offers $75$\kms\ velocity resolution and 1\arcsec\ spatial
resolution, which allows a determination of the relative positions of
the OH masers and the shock--heated gas. We detected \h\ emission
towards all four OH/SNR sources we observed (G359.1-0.5, G349.7+0.2,
G357.7-0.1, G337.0-0.1). Shock excitation of \h\ is inferred either
from the ratio of two \h\ 2--1/1--0 S(1) lines, or from the
correlation between \h\ and synchrotron radio emission. \h\
observations towards G349.7+0.2  are presented in Figure~\ref{fig-candy}.

\section{OH absorption}

Theoretical models do not predict OH in any great abundance, because
for temperatures above 400 K any OH formed from O and \h\ will be
rapidly converted into H$_2$O. However, X--rays from the SNR interior
can penetrate the surrounding molecular cloud, photo-ejecting electrons
which collisionaly excite \h. The subsequent radiative decay of \h\
contributes to a secondary UV radiation field, which converts part of
 H$_2$O back to OH after the temperature drops below \mbox{200 K} (Wardle 1999). 

The OH column necessary to produce \oh masers can be
observed in absorption at 18 cm against the background SNR radio
continuum. We derived OH column density of $\sim$\e{16}\cm{-2} towards
three OH/SNR objects (W28, \candy, G337.0-0.1) we observed using the
Australia Telescope Compact Array (ATCA), which is consistent with the
model for \oh maser production. In Figure~\ref{fig-candy} we present
results on \candy. The maximum OH absorption occurs towards the OH
masers  and is well correlated with shocked \h\ emission. This is a
clear indication that the enhancement of the OH gas is associated with shocked regions rather than background continuum intensity, as is usually found.

\acknowledgments JSL was supported by the Australian Government
International Postgraduate Research Scholarship. MW was supported by
the Australian Research Council.  The Australia Telescope is funded by
the Commonwealth of Australia for operation as a National Facility
managed by CSIRO. SEST is operated by the Swedish National
Facility for Radio Astronomy, Onsala Space Observatory and by the
European Southern Observatory (ESO).

\end{document}